\documentstyle[aps,twocolumn,epsfig]{revtex}

\begin{document}

\title{Double-Occupancy Errors, Adiabaticity, and Entanglement\\ of Spin-Qubits
in Quantum Dots}

\author{John Schliemann$^{1,2}$, Daniel Loss$^{3}$, and A. H. MacDonald$^{1,2}$}

\address{$^1$Department of Physics, Indiana University, 
Bloomington, IN 47405\\
         $^2$Department of Physics, The University of Texas, Austin, TX 78712\\
$^3$Department of Physics and Astronomy, University of Basel,
Klingelbergstra{\ss}e 82,
CH-4056 Basel, Switzerland}

\date{August 2000}

\maketitle

\begin{abstract} 

Quantum gates that temporarily increase singlet-triplet
splitting in order to swap electronic spins in coupled quantum dots, 
lead inevitably to a finite double-occupancy probability for both
dots.  By solving the time-dependent Schr\"odinger equation for a coupled
dot model, we demonstrate that this does not necessarily lead to 
quantum computation errors.   Instead, the coupled dot ground state
evolves quasi-adiabatically for typical system parameters so that 
the double-occupancy probability at the completion of 
swapping is negligibly small.
We introduce a measure of entanglement which explicitly takes 
into account the possibilty of double occupancies and provides a 
necessary and sufficient criterion for entangled states.
\end{abstract}

%%%%%%%%%%%%%%%%%%%%%%%%%%%%%%%%%%%%%%%%%%%%%%%%%%%%%%%%%%%%%%%%%%%%%%

\section{Introduction}

In the past several years there has been a great deal of interest in possible
physical realizations of quantum computing bits and operations\cite{Steane}.
Among the  various proposals, solid state systems are particularly attractive
since they are more easily integrated into large quantum networks.
In particular, semiconductor 
nanostructures which use the spin degree of freedom of the electrons\cite{LoDi:98}
(rather than their charge) for information processing are of  special
interest since they can take advantage of the comparatively 
long spin coherence times in such materials \cite{KiAw,KiAw2,Gupta}. 

A key challenge is the construction of systems composed of 
two coupled quantum dots which can be coupled to perform
swap operations ${\cal U}_{SW}$, i.e. unitary two-qubit operations  
which interchange the spin states (qubits) of the
electrons on the two dots \cite{LoDi:98,BLD:99,BSL:99,Engel,HuDa:99,DFXZ:99}. 
By combining the ``square root'' ${\cal U}_{SW}^{1/2}$ 
of such a swap with other isolated-qubit manipulations one can construct a quantum
XOR gate. A quantum XOR gate, along with isolated-qubit operations, has been
shown to be sufficient for the implementation of any quantum algorithm
\cite{DiV:95}. Hence
a practical and reliable realization of a swap gate would be an important
step towards the fabrication of a solid state quantum computer.

The swap operation of electron spin states in a double dot system can 
be realized in principle by turning on a time-dependent
exchange coupling between the spins as a ``source of 
entanglement''.  
In practice the exchange interaction is provided by 
singlet-triplet splitting in a double-dot, which is always
accompanied by a finite inter-dot electron tunneling amplitude.

In a recent work Das Sarma and coworkers \cite{HuDa:99,DFXZ:99}
emphasized
 that exchange interactions in the range of interest are 
accompanied by 
a substantial probability, during the swap operation, that both electrons 
will be 
on the same dot.  In this paper we demonstrate that, contrary to naive 
expectations,
these virtual double-occupancies will {\em not}, under circumstances typically
envisioned, lead to an important increase in quantum computing errors.
{\em Double occupancy is not a fatal problem for quantum dot based quantum
computing with spins.}  The occurrence of double occupancies {\em during} the
swapping  process does
not lead to processing errors, provided that the double occupancies are 
sufficiently
suppressed when the swapping of spin states is completed.  The principle
purpose of the present paper is to illustrate this basic feature within the
Hund-Mulliken description of a quantum dot hydrogen molecule. We will see that,
in a system of identical dots,
the time evolution of this system can be reduced to the problem of a 
pseudospin-$\frac{1}{2}$ in a time-dependent pseudomagnetic field. In 
particular,
the question of whether double occupancies are a severe obstacle for swap
operations in the quantum dot system is equivalent to the question of how
close the pseudospin dynamics is to its adiabatic limit.
Simple numerical studies presented in Sec. IV show that the 
pseudospin has an approximately adiabatic time evolution for
a ramarkably broad range of coupling ramp times. It turns out
that this behavior holds even if the inversion symmetry
along the x-axis connecting the dots is broken (e.g. in the presence
of an electric field).

A secondary purpose of this work is to introduce a coordinate-independent 
measure
of entanglement appropriate for the Hilbert space of the above system.
This quantity provides a necessary and sufficient criterion for the
entanglement of quantum states. It differs from other entanglement criteria
proposed in the literature \cite{entcrit}
in so far as it takes into account states with
double occupancies. This generalizes the typical situation of
Einstein-Podolsky-Rosen experiments.  We expect this measure of 
entanglement to be useful in the theoretical study of coupled quantum dots
(or similar quantum-confined nanostructures), independent of the particular model
considered here.

%%%%%%%%%%%%%%%%%%%%%%%%%%%%%%%%%%%%%%%%%%%%%%%%%%%%%%%%%%%%%%%%%%%%%%

\section{The model}
\label{model}

We consider a system of two electrons in two laterally coupled quantum dots.
The experimental motivation for the model described below has been discussed
elsewhere \cite{BLD:99,BSL:99,Engel}. Here we just summarize its basic features.

The Hamiltonian is given by ${\cal H}=T+C$, where $C$ denotes the Coulomb
repulsion between the electrons, and $T=\sum_{i=1,2}h_{i}$ is the 
one-particle part with
\begin{equation}
h_{i}=\frac{1}{2m}\left(\vec p_{i}+\frac{e}{c}\vec A(\vec r_{i})\right)^{2}
+V(\vec r_{i})\,.
\label{opham}
\end{equation}
The one-particle Hamiltonian $h_{i}$ describes electron dynamics confined
to the $xy$-plane in a perpendicular magnetc field B. The effective mass 
$m$ is a material-dependent parameter.
The coupling of the dots (which includes tunneling)
is modeled by a quartic potential 
\begin{equation}
V(x,y)=\frac{m\omega^{2}_{0}}{2}
\left(\frac{1}{4a^2}\left(x^{2}-a^{2}\right)^{2} +y^{2}\right)\,,
\end{equation}
which separates into two harmonic wells of frequency $\omega_{0}$ (one for
each dot) in the limit $2a\gg 2a_{0}$, where $a$ is half the distance between 
the dots and $a_{0}=\sqrt{\hbar/m\omega_{0}}$ is the effective Bohr radius
of a dot. 

Following Burkard {\it et al.} \cite{BLD:99} we employ the Hund-Mulliken
method of molecular orbits to describe the low-lying spectrum of our system.
This approach concentrates on the lowest orbital states in each dot and is
an extension of the Heitler-London method also discussed in \cite{BLD:99}.
The Hund-Mulliken approach accounts for
double occupancies and is therefore suited to investigate the questions 
at issue here.

In the usual symmetric gauge $\vec A=B(-y,x,0)/2$ the Fock-Darwin ground state
of a single dot with harmonic confinement centered around $\vec r=(\pm a,0,0)$
reads
\begin{eqnarray}
\varphi_{\pm a}(x,y) & = & \sqrt{\frac{m\omega}{\pi\hbar}}
\exp\left(\frac{m\omega}{2\hbar}\left(\left(x\mp a\right)^{2}+y^{2}\right)
\right)\nonumber\\
 & & \cdot\exp\left(\pm\frac{i}{2}y\frac{a}{l_{B}^{2}}\right)\,,
\label{dotstates}
\end{eqnarray}
where $l_{B}=\sqrt{\hbar c/eB}$ is the magnetic length, and the frequency 
$\omega$ is given by $\omega^{2}=\omega_{0}^{2}+\omega_{L}^{2}$ where
$\omega_{L}=eB/2mc$ is the usual Larmor frequency. From these non-orthogonal
one-particle states we construct the orthonormalized states $|A\rangle$ and 
$|B\rangle$ with wavefunctions
\begin{eqnarray}
\langle\vec r|A\rangle & = &
\frac{1}{\sqrt{1-2Sg-g^{2}}}\left(\varphi_{+a}-g\varphi_{-a}\right)\,,\\
\langle\vec r|B\rangle & = &
\frac{1}{\sqrt{1-2Sg-g^{2}}}\left(\varphi_{-a}-g\varphi_{+a}\right)\,,
\end{eqnarray}
with $S$ being the overlap between the states (\ref{dotstates}) and
$g=(1-\sqrt{1-S^{2}})/S$. 
For appropriate values of system parameters such as the interdot distance
and the external magnetic field, the overlap $S$ becomes exponentially small
\cite{BLD:99}. In this limit an electron in one of the states
$|A\rangle$, $|B\rangle$ is predominantly localized around
$\vec r=(\pm a,0,0)$.
In the following we consider this case and use these states as basis states 
to define qubits, i.e.
qubits are realized by the spin state of an electron in either orbital 
$|A\rangle$, or orbital $|B\rangle$. 

An appropriate basis set for the six-dimensional two-particle Hilbert space is
given (using standard notation) by the three spin singlets
\begin{eqnarray}
|S_{1}\rangle & = & \frac{1}{\sqrt{2}}
\left(c^{+}_{A\uparrow}c^{+}_{B\downarrow}-
c^{+}_{A\downarrow}c^{+}_{B\uparrow}\right)|0\rangle\,,\\
|S_{2}\rangle & = & \frac{1}{\sqrt{2}}
\left(c^{+}_{A\uparrow}c^{+}_{A\downarrow}+
c^{+}_{B\uparrow}c^{+}_{B\downarrow}\right)|0\rangle\,,\\
|S_{3}\rangle & = & \frac{1}{\sqrt{2}}
\left(c^{+}_{A\uparrow}c^{+}_{A\downarrow}-
c^{+}_{B\uparrow}c^{+}_{B\downarrow}\right)|0\rangle\,,
\end{eqnarray}
and the triplet multiplet, 
\begin{eqnarray}
|T^{-1}\rangle & = &
c^{+}_{A\downarrow}c^{+}_{B\downarrow}|0\rangle\,,\\
|T^{0}\rangle & = & \frac{1}{\sqrt{2}}
\left(c^{+}_{A\uparrow}c^{+}_{B\downarrow}+
c^{+}_{A\downarrow}c^{+}_{B\uparrow}\right)|0\rangle\,,\\
|T^{1}\rangle & = & c^{+}_{A\uparrow}c^{+}_{B\uparrow}|0\rangle\,.
\end{eqnarray}
The three triplet states are  degenerate (typically we can ignore possible
Zeeman splittings\cite{BLD:99}) and have the
common  eigenvalue, 
\begin{equation}
\varepsilon_{T}=2\varepsilon+V_{-}\,,
\end{equation}
where we have defined 
\begin{equation}
\varepsilon=\langle A|h|A\rangle=\langle B|h|B\rangle
\end{equation}
and
\begin{equation}
V_{-}=\langle T^{\alpha}|C|T^{\alpha}\rangle\quad,\quad
V_{+}=\langle S_{1}|C|S_{1}\rangle\,.
\end{equation}

An important further observation is that, as a consequence of inversion 
symmetry along the axis connecting the dots,
the Hamiltonian does not
have any non-zero matrix elements between the singlet state $|S_{3}\rangle$
and other states. Hence, $|S_{3}\rangle$ is, independently of
the system parameters, an eigenstate. The eigenvalues of the triplet and
$|S_{3}\rangle$, however, do depend on system parameters. The
Hamiltonian acting on the remaining space spanned by $|S_{1}\rangle$ and
$|S_{2}\rangle$ can be written as
\begin{equation}
{\cal H}=2\varepsilon+\frac{1}{2}U_{H}+V_{+}-\left(
\begin{array}{cc}
 U_{H}/2 & 2t_{H} \\ 2t_{H} & -U_{H}/2
\end{array}\right)
\label{2ham}
\end{equation}
where 
\begin{equation}
t_{H}=\langle A|h|B\rangle=\langle B|h|A\rangle
\end{equation}
and
\begin{equation}
U_{H}=\langle S_{2}|C|S_{2}\rangle-V_{+}\,.
\end{equation}
The nontrivial part of (\ref{2ham}) is a simple Hubbard Hamiltonian
and can be identified as the Hamiltonian of a 
pseudospin-$\frac{1}{2}$-object in a
pseudomagnetic field having a component $U_{H}$ in the $\hat z$-direction and 
$4t_{H}$ in 
the $\hat x$-direction of pseudospin space.  [Note that this 
pseudospin is not related to the spin degree of freedom which provides the 
qubit!]
The space spanned by $|S_{1}\rangle$ and 
$|S_{2}\rangle$ contains the ground state of the system.
The basis states 
themselves are eigenstates only in the case of vanishing tunneling amplitude
$t_{H}$ where $|S_{1}\rangle$ is the ground state.
In all other cases, the ground state has an admixture of double
occupied states contained in $|S_{2}\rangle$. 
The energy gap between the triplet and the singlet ground state is
\begin{equation}
\varepsilon_{T}-\varepsilon_{S0}=V_{+}-V_{-}-\frac{U_{H}}{2}+
\frac{1}{2}\sqrt{U^{2}_{H}+16t^{2}_{H}}\,.
\end{equation}

A swap operation in the present system is a unitary transformation which turns
a state having the qubits in different states, say,
\begin{equation}
c^{+}_{A\uparrow}c^{+}_{B\downarrow}|0\rangle=
\frac{1}{\sqrt{2}}\left(|T^{0}\rangle+|S_{1}\rangle\right)\,,
\label{instate}
\end{equation}
into a state where the contents of the qubits is interchanged,
\begin{equation}
c^{+}_{A\downarrow}c^{+}_{B\uparrow}|0\rangle=
\frac{1}{\sqrt{2}}\left(|T^{0}\rangle-|S_{1}\rangle\right)\,.
\label{outstate}
\end{equation}
These two states are eigenstates in the case $V_{+}=V_{-}$ and $t_{H}=0$
for which the singlet-triplet splitting vanishes.

As discussed in references \cite{LoDi:98,BLD:99}, swapping may be achieved 
by the action of a gate that lowers the potential barrier between the 
quantum dots.  This
leads to exponentially larger values for both $V_{+}-V_{-}$ and 
$t_H$.  It is adequate for our purposes to consider a model where 
$V_{+}=V_{-}$ (consistent with the above limit of small overlap $S$), and 
the singlet-triplet splitting results entirely from 
$t_H$.  If the duration and amplitude of a tunneling pulse
is adjusted
appropriately, the relative phase between the singlet and the triplet state 
involved picks up a shift of $\pi$, and a swapping operation is performed.

As pointed out in \cite{DFXZ:99} a finite tunneling 
amplitude necessarily leads to a finite probability for double occupancies of 
qubit states. If double occupancy errors occurs to any sizable extent 
{\em as a result} of the
swapping process, any quantum computation based on this hardware is likely to 
fail. 
However, if the double occupancies are sufficiently 
rare {\em after} the swapping process, errors in the quantum computation can 
likely
be corrected dynamically.
An important observation is that the double-occupancy probability 
{\em after} the swap vanishes in the adiabatic limit, {i.e.} if the 
ramp time $\tau$ of the quantum gate is such that 
$\hbar /\tau $ is much larger than the pseudospin splitting 
$\sqrt{U_H+16t_H^2}$. 
This follows since the non-adiabatic effects can arise only from 
the states $|S_{1}\rangle$ and $|S_{2}\rangle$, which have a non-trivial
time evolution when the tunneling amplitude $t_{H}$ is time-dependent. 
Thus, the question
of whether double occupancies are problematic for swap operations in the
present system is reduced to the question of how close the motion of
a spin-$\frac{1}{2}$-object in a time-dependent magnetic field is to its
adiabatic limit. This will be investigated further in Sec. \ref{numres}. 

The reduction of the dynamics to the time evolution of a two-level-system
relies on the fact that the system has inversion symmetry along the
$\hat x$-axis in real space connecting the dots. This symmetry can be
broken if odd powers of the particle coordinates $x_{i}$ are added to
the Hamiltonian (\ref{opham}) like for example the potential of a
homogeneous electric field. However, the only addional matrix element
due to such terms in the Hamiltonian occurs in the subspace of double
occupied states between the singlets $|S_{2}\rangle$ and
$|S_{3}\rangle$. Thus, in the presence of an electric field 
${\cal E}=-eE\sum_{i}x_{i}$ the Hamitonian acting on the singlet
subspace spanned by $|S_{1}\rangle$, $|S_{2}\rangle$, $|S_{3}\rangle$ reads
\begin{eqnarray}
{\cal H} & = & 2\varepsilon+\frac{1}{2}U_{H}+V_{+}\nonumber\\
 & & -\left(
\begin{array}{ccc}
 U_{H}/2 & 2t_{H}  & 0 \\ 
 2t_{H} & -U_{H}/2 & F \\
 0 & F & -U_{H}/2+2X
\end{array}\right)
\label{3ham}
\end{eqnarray}
with the real matrix element $F=\langle S_{2}|{\cal E}|S_{3}\rangle$ and
\begin{equation}
2X=\langle S_{2}|C|S_{2}\rangle-\langle S_{3}|C|S_{3}\rangle
=2\langle A|\langle A|C|B\rangle|B\rangle\,.
\end{equation}
With a finite matrix element $F$ the dynamics of the system  
is slightly more complicated, but also in this case the only coupling
of the two-qubit states (\ref{instate}) and (\ref{outstate}) to the
subspace of double occupied states is
provided by the tunneling amplitude $t_{H}$. Therefore, with respect
to the adiabaticity of the swapping process, the situation can be
expected to be not very different from the one with inversion symmetry
between the dots. This will be verified in Sec. {\ref{numres}.

So far we have not considered a possible Zeeman coupling to the electron spin.
This would not change the situation essentially since all states involved in the
swapping  process ($|T^{0}\rangle$, $|S_{1}\rangle$, $|S_{2}\rangle$, and
eventually $|S_{3}\rangle$) have the
total spin quantum number $S^{z}=0$.

%%%%%%%%%%%%%%%%%%%%%%%%%%%%%%%%%%%%%%%%%%%%%%%%%%%%%%%%%%%%%%%%%%%%%%

\section{Entangled states}
\label{entsta}

Before analysing further the possibility of performing swap operations in the 
above system, let us introduce an appropriate measure for the entanglement of 
its quantum states. Consider a system of two fermions living in a 
four-dimensional
one-particle space. A general state vector in this six-dimensional Hilbert
space can be written as
\begin{equation}
|w\rangle=w_{ab}\,c^{+}_{a}c^{+}_{b}|0\rangle
\label{defstate}
\end{equation}
where a summation convention is understood for repeated latin indices
$a,b,\dots\in\{1,2,3,4\}$ running over the orthonormalised one-particle 
states. The coefficient matrix $w$ can be assumed to be
antisymmetric, $w_{ab}=-w_{ba}$. The normalisation condition reads
\begin{equation}
\langle w|w\rangle=1\quad\Leftrightarrow\quad
{\rm Tr}(\bar w\,w)=-\frac{1}{2}\,,
\end{equation}
where the bar denotes complex conjugation. A two-particle state of the
form (\ref{defstate}) is in general entangled, i.e. cannot be written
as a single Slater determinant. $|w\rangle$ is non-entangled, i.e. a single
Slater determinant, if $w$ has the form
\begin{equation}
w_{ab}=\frac{1}{2}\left(z^{1}_{a}z^{2}_{b}-z^{1}_{b}z^{2}_{a}\right)
\label{slater}
\end{equation} 
for two orthonormal spinors $z^{1}$ and $z^{2}$,
$z^{i}_{a}\bar z^{j}_{a}=\delta^{ij}$. We note that for a given
non-entangled state $|w\rangle$ the choice of spinors $z_{1}$ and
$z_{2}$ is by no means unique since any SU(2) transformation among
these two occupied one-particle states leads to the same two-particle
state vector $|w\rangle$. Hence, for a given non-entangled state $|w\rangle$
there is a three-dimensional manifold of spinors fulfilling equation
(\ref{slater}).

We define the {\it dual matrix} $\tilde w$ of $w$ by
\begin{equation}
\tilde w_{ab}=\frac{1}{2}\varepsilon^{abcd}\bar w_{cd}
\label{defdual}
\end{equation}
with $\varepsilon^{abcd}$ denoting the totally antisymmetric unit tensor in
four dimensions.
The scalar product of a state $|w\rangle$ with its dual state
$|\tilde w\rangle$ can be written as
\begin{equation}
\langle\tilde w|w\rangle=\varepsilon^{abcd}w_{ab}w_{cd}
=8\left(w_{12}w_{34}+w_{13}w_{42}+w_{14}w_{23}\right)\,.
\end{equation}
This cyclic sum vanishs identically if $w$ has the form (\ref{slater}).
Hence the vanishing of
\begin{equation}
\eta(w):=|\langle\tilde w|w\rangle|
\label{defeta}
\end{equation}
is a necessary condition for $|w\rangle$ being a single Slater determinant.
Moreover, in the Appendix it is shown that $\eta(w)=0$ is actually
also a suffcient condition for $|w\rangle$ being non-entangled. Thus, $\eta$
defines a measure of entanglement which is exactly zero for non-entangled
states.   $\eta(w)\ne0$ is therefore a necessary and sufficient condition for
entanglement of quantum states. Maximally entangled states are characterized
by the fact that they are collinear with their dual states, $\eta(w)=1$.
As simple examples, we consider the basis states used in the preceding 
section:
the states $|T^{-1}\rangle$ and $|T^{1}\rangle$ are single Slater
determinants, while all other basis states are maximally entangled.

The matrix $w$ transforms under a unitary transformation of the
one-particle space,
\begin{equation}
c^{+}_{a}\mapsto{\cal U}c^{+}_{a}{\cal U}^{+}=U_{ba}c^{+}_{b}\,,
\end{equation}
as 
\begin{equation}
w\mapsto UwU^{T}\,,
\end{equation}
where $U^{T}$ is the transpose (not the adjoint) of $U$. It is straightforward
to see that $\eta$ is invariant under such transformations, and the 
determinant
of $w$ remains the same up to a possible phase factor. Thus, the 
entanglement of a state $|w\rangle$ quantified by $\eta$ does not depend on
the basis chosen for the one-particle space, which is of course a
necessary requirement for a measure of entanglement.

The dualisation of a state can be identified as a particle-hole-transformation,
\begin{equation}
{\cal U}_{p-h}c^{+}_{a}{\cal U}^{+}_{p-h}=c_{a}\quad,\quad
{\cal U}_{p-h}|0\rangle=c^{+}_{1}c^{+}_{2}c^{+}_{3}c^{+}_{4}|0\rangle\,,
\label{ph1}
\end{equation}
followed by complex conjugation. In fact, 
\begin{equation}
{\cal U}_{p-h}|w\rangle=-|\bar{\tilde w}\rangle\,.
\label{ph2}
\end{equation}
We note that the complex conjugations in (\ref{defdual}) and (\ref{ph2}) are
unimportant for states $|w\rangle$ such that $\eta(w)=0$,
since a single Slater determinant is always orthogonal to its particle-hole
conjugate, irrespective of a possible phase transformation
of a prefactor. However,
the complex conjugation in the definition (\ref{defdual}) is essential to  
the sufficiency of the above condition. As an example consider a state
$|w\rangle$ with $w_{12}=w_{34}=1/4$, $w_{13}=w_{24}=i/4$ and 
$w_{14}=w_{23}=0$. This is
clearly a maximally entangled state, $\eta(w)=1$, while its scalar product
with the complex conjugate of its dual state is 
$\langle\bar{\tilde w}|w\rangle=0$. 

We also mention the following identity for the determinant of $w$:
\begin{equation}
\det w=\left(\frac{1}{8}\langle\tilde w|w\rangle\right)^{2}
\label{det}
\end{equation}
Hence it follows that also $|\det w|$ could be used as a measure of 
entaglement.
Equation (\ref{det}) is important for the proof of the sufficiency
of our criterion for non-entangled states, as explained in detail in the 
Appendix.

A convenient choice to make contact between the general state labels
$a,b,\dots\in\{1,2,3,4\}$ used here and the basis states of the
preceding section is given by $(1,2,3,4)=
(A\uparrow,A\downarrow,B\uparrow,B\downarrow)$.
With this convention, a state vector spanned by $|S_{2}\rangle$ and
$|S_{3}\rangle$ only has $w_{12}$ and $w_{34}$ as its only independent
non-zero coefficients in $w$. Such a state lies fully in the subspace
of double occupancies, and its entanglement is purely due to the
orbital degrees of freedom:
\begin{equation}
\eta_{orb}=8|w_{12}w_{34}|\,.
\label{orbit}
\end{equation}
On the other hand, a state spanned by $|S_{1}\rangle$ and $|T^{0}\rangle$
has no double occupancies and 
is entangled purely with respect to the spin degrees of freedom,
\begin{equation}
\eta_{spin}=8|w_{14}w_{23}|\,.
\label{spin}
\end{equation}
For a general state vector, both kinds of entanglement (orbital and
spin) contribute to
$\eta(w)$.

%%%%%%%%%%%%%%%%%%%%%%%%%%%%%%%%%%%%%%%%%%%%%%%%%%%%%%%%%%%%%%%%%%%%%%

\section{Results for the swapping process}
\label{numres}

We now continue with our investigation of the dynamics of the double
quantum 
dot qubit swapping process generated by a time-dependent tunneling
amplitude. 

Let us first consider the case of inversion symmetry
along the axis connecting the dots.
As explained in Sec. \ref{model} this problem 
can be reduced essentially to the time evolution of a  
pseudospin-$\frac{1}{2}$-object in a magnetic field having a time-dependent
component in the $x$-direction of the pseudospin space. In the course of swapping,
the triplet contribution to the incoming state (\ref{instate}) will 
just pick up a phase factor according to its constant eigenvalue, while the
singlet contribution will mix with the other singlet $|S_{2}\rangle$.
Therefore, a finite probability for double occupancies will necessarily
arise {\em during} the swapping process. However, if these amplitudes can be
suppressed sufficiently when the swapping is complete (as in
the adiabatic limit), errors in the quantum computation
can be avoided. Thus we are left with the question of how close the dynamics
of our formal spin-$\frac{1}{2}$-object is to its adiabatic limit.
We note that, in the adiabatic limit, no Berry phase
occurs in the time evolution of the singlet states,
since the motion of the formal spin is restricted to a plane. Hence 
the solid angle encircled in a round trip is strictly zero. 

The integration of the Schr\"odinger equation for our time-dependent two-level
problem is in general non-elementary. However, there is a considerable body of
literature, starting with early work by Landau \cite{Lan:32}, Zener 
\cite{Zen:32}, 
and Rosen and Zener \cite{RoZe:32}, where particular cases of this
problem were reduced to 
well-known differential equation of mathematical physics such as the 
hypergeometric equation. 
This work was reviewed and generalized very recently in 
\cite{Ish:00}. However, such an approach still works only for special
time-dependent Hamiltonians, i.e., in the present context, only for special 
shapes of
the tunneling pulse $t_{H}(t)$, and many quantities of 
interest are given by complicated non-elementary
expressions which require numerical evaluation. For this reasons, and for the 
sake of brevity of our presentation, we shall resort to numerical integrations
of the Schr\"odinger equation. From such studies we will see that the
range of adiabaticity is remarkably large. 
Our numerical findings will be corroborated and made physically plausible
by well-known analytical results for Landau-Zener-type transitions in
simplified cases. 

To be specific, we consider a time-dependent tunneling of the form
\begin{equation}
t_{H}(t)=\frac{\Delta}
{1+\frac{\cosh\left(t/\tau\right)}{\cosh\left(T/(2\tau)\right)}}\,.
\label{tunnel}
\end{equation}
This is a tunneling pulse which is switched on and off exponentially with
a characteristic time $\tau$. It has a duration of $T$ and an amplitude
given by $\Delta$ (for $T\gg\tau$). Therefore this form is flexible enough 
to describe the essential features of a pulse.
The exponential switching is motivated
by the exponential-like dependence of the tunneling matrix element on external
parameters \cite{BLD:99}.

A typical situation is shown in figure \ref{fig1} for a switching time
of $\tau=4\hbar/U_{H}$, an amplitude of $\Delta=U_{H}/8$ and the duration
$T$ adjusted to enable  single swap operation. The figure shows the results
of a numerical integration of the time-dependent Schr\"odinger equation
using the fourth order Runge-Kutta scheme. The time-dependent
tunneling amplitude $t_{H}(t)$ is plotted (in units $U_{H}$) as a dotted line.
The square amplitude of
the incoming state (\ref{instate}) and the outgoing state (\ref{outstate}) are
shown as thick lines. The square amplitudes of the singlets 
$|S_{1}\rangle$ and $|S_{2}\rangle$ are denoted by $|\varphi_{1}|^{2}$ and
$|\varphi_{2}|^{2}$, respectively, and plotted as long-dashed lines.
The probability of double occupancies is given by $|\varphi_{2}|^{2}$.
As one can see from the figure, this quanity is finite during the
swapping process but strongly suppressed afterwards. The measure of
entanglement $\eta(t)$ is also shown in the figure. It is zero for the 
non-entangled incoming and outgoing state, and achieves its 
maximum value of almost unity in the middle of the process. This quantifies and
shows explicitly the entanglement of the quantum state during the
swapping process. 

The probability $|\varphi_{2}|^{2}$ for double occupancy after switching off
the tunneling depends on the switching time $\tau$, the amplitude $\Delta$ and 
also on the duration $T$ of the tunneling pulse, i.e. on the exact time when
the switching off sets in. However, our numerics suggest that there is 
an upper bound for $|\varphi_{2}|^{2}$ at given $\tau$ and $\Delta$.
In the above example the double occupancy probability after the swapping
process is smaller than $10^{-10}$, which is a very tiny value.
A typical order of 
magnitude for the double occupancy probability is $10^{-6}$ for amplitudes
$\Delta<U_{H}$ and switching times $\tau>4\hbar/U_{H}$. In fact, also
larger
values of $\Delta$ (being still comparable with $U_{H}$) can be possible, 
leading to double occupancy probabilities of the same order,
while this probability significantly increases if $\tau$ 
becomes smaller than $4\hbar/U_{H}$. Thus, this value characterizes the 
region where the motion of the system is close to its adiabatic limit and
is remarkably small on the natural time scale of the system given by
$\hbar/U_{H}$, while adiabatic behavior is in general expected for a
particularly slow time evolution.

This large range of quasi-adiabatic behavior can be understood qualitatively
by considering a simplified situation where the tunneling is switched on and 
off linearly in time and is constant otherwise. Then, non-adiabtic effects
can occur only during the sharply defined switching processes. For simplicity,
we consider the first switching process only where the tunneling has
the time dependence $t_{H}=(\Delta/\tau)t$, $t\in[0,\tau]$. To enable 
analytical progress let us further assume $t\in[-\infty,\infty]$, which should 
lead to an upper bound for the probability of non-adiabatic transitions
due to the switching. This problem was considered a long time ago
by Landau \cite{Lan:32} and by Zener \cite{Zen:32}. 
The result of reference \cite{Zen:32} for the probability of non-adiabatic 
transitions reads 
\begin{equation}
P_{{\rm nad}}=e^{-\alpha}
\label{traprob}
\end{equation}
with an adiabaticity parameter 
\begin{equation}
\alpha=\frac{\pi}{8}\frac{U_{H}^{2}}{\hbar(\Delta/\tau)}\,.
\label{adpar}
\end{equation}
We see
that the probability for non-adiabatic transitions is exponentially suppressed
with increasing switching time $\tau$. This exponential dependence explains
qualitatively the above observation of a large range of quasi-adiabatic
bahavior. To obtain an estimate for a nonlinear switching one may replace
the ratio $(\Delta/\tau)$ in the denominator of (\ref{adpar}) by the maximum
time derivative of the tunneling $t_{H}(t)$ (giving
$\alpha=\pi U_{H}^{2}/3\hbar(\Delta/\tau)$ for the pulse (\ref{tunnel})).

A similar exponential dependence of the  probability for 
non-adiabatic transitions on the switching time $\tau$ was also found 
analytically by Rosen and Zener \cite{RoZe:32} for a particular 
two-parametric pulse of the form 
\begin{equation}
t_{H}(t)=\Delta/\cosh(t/\tau)\,.
\end{equation}
In this case non-adiabatic transitions occur with a probability
\begin{equation}
P_{{\rm nad}}=\sin^{2}\left(\Delta\tau/(2\hbar)\right)/
\cosh^{2}\left(U_{H}\tau/(2\hbar)\right)\,.
\label{traprob2}
\end{equation}
To illustrate the behavior in the strongly non-adiabatic case we have plotted
in figure \ref{fig2} $|\varphi_{1}|^{2}$ and
$|\varphi_{2}|^{2}$ for the same situation as in figure \ref{fig1} but with a 
four times smaller 
ramp time of only $\tau=\hbar/U_{H}$. In this case small oscillations occur in
the time-evolution of these two quantities during the tunneling pulse, 
which can be understood in terms
of the eigenspectrum at a given tunneling $t_{H}=\Delta$. Additionally, a
sizable double occupancy probability of about $0.005$ remains after the pulse,
as shown in the inset.

Figure \ref{fig3} shows a square root of a swap, which is obtained from the
situation of figure \ref{fig1} by halfing the duration $T$ of the tunneling 
pulse. The resulting state is a fully entangled complex 
linear combination of the
states $|S_{1}\rangle$ and $|T^{0}\rangle$, or, equivalently, of the 
incoming state (\ref{instate}) and the outgoing state (\ref{outstate}) of the
full swap. Again, the weight of the
doubly occupied state $|S_{2}\rangle$ is strongly suppressed after
the tunneling pulse. As a consequence, Eq. (\ref{orbit}) implies that
$\eta_{orb}=0$ after completion of switching, while
$\eta=\eta_{spin}=8|w_{14}w_{23}|=1$. This shows that the entanglement
of the two electrons is entirely in the spin (and not in the orbital)
degrees of freedom after switching.

Let us finally consider swapping processes when the inversion symmetry along
the axis connecting the dots is broken. Such processes are governed by the 
Hamiltonian (\ref{3ham}) in the presence of a finite matrix element $F$.
Our numerical results are in this case qualitatively the same as before
with the admissible switching times $\tau$ slightly growing with increasing
$F$. In figure \ref{fig4} we illustrate our findings for a comparatively large
off-diagonal element $F=0.4U_{H}$. The additional Coulomb matrix element is
$X=0.2U_{H}$, and the parameters of the tunneling pulse are
$\tau=8\hbar/U_{H}$ and $\Delta=U_{H}/8$ with a duration $T$ appropriate
for a single swapping. As a result, a clean swapping operation can be
performed also in the absence of inversion symmetry.

%%%%%%%%%%%%%%%%%%%%%%%%%%%%%%%%%%%%%%%%%%%%%%%%%%%%%%%%%%%%%%%%%%%%%%%%%

We note that the Hund-Mulliken scheme used here is restricted to the
low-energy  sector where only the lowest single-particle energy
levels (with typical spacings $\delta\epsilon$) are kept. 
For this scheme to be valid also in a switching process, we need to require
that time-dependent changes must  be performed adiabatically also with respect to
the time scale set by
$\hbar/\delta\epsilon$, i.e. we need $\tau >
\hbar/\delta\epsilon$\cite{BLD:99}. On the other hand, 
to suppress double occupancy errors we have seen that 
the adiabaticity parameter $\alpha$ of Eq.
(\ref{adpar}) must exceed one, implying that 
$\tau > 8\hbar\Delta/ (\pi U_H^2)$.
Thus, the adiabaticity condition for switching becomes more generally,
\begin{equation}
\tau> \tau_{min}:= \max
\left\{ {\hbar\over \delta\epsilon}, {8\hbar\over \pi}
{\Delta
\over U_H^2}\right\}\, .
\end{equation}
There are now two particular cases we can distinguish. First, if the
effective Coulomb charging energy exceeds the level spacing,
i.e. $U_H>\delta\epsilon$,
we obtain $\tau_{min}=\hbar/\delta\epsilon$, 
since for consistency we have $\Delta <\delta\epsilon$. Thus, when the
switching is adiabatic with respect to the scale set by $\delta\epsilon$,
errors due to double occupancy are automatically excluded.
In the second case with $U_H < \sqrt{\Delta \delta\epsilon}<\delta\epsilon$
(``ultrasmall quantum dots"),
we obtain $\tau_{min}=8\hbar\Delta/(\pi U_H^2)$, which means that the
overall  condition for adiabaticity is determined by the no-double occupancy
criterion.  Using typical material parameters for GaAs quantum
dots\cite{kouwenhoven}, we can estimate\cite{BLD:99} that $\tau_{min}$ is of 
the order of 50 ps.

%%%%%%%%%%%%%%%%%%%%%%%%%%%%%%%%%%%%%%%%%%%%%%%%%%%%%%%%%%%%%%%%%%%%%%%%%

\section{Conclusions}
\label{conclu}

We have  studied a double quantum dot system as a quantum gate swapping the 
electronic spin states on the two dots. Within
Hund-Mulliken approach, the dynamics of such a system having inversion
symmetry along the axis connecting the dots 
reduces to the problem of a pseudospin-$\frac{1}{2}$-object in a 
time-dependent pseudomagnetic field. By solving the time-dependent Schr\"odinger
equation numerically we demonstrate the possibility of performing swap 
operations and investigate the role of double ocupancies of the dots.
These double occupancies are found to be (exponentially) strongly reduced, as a
result of the swapping process, for a large range of system parameters and are
therefore not  a principle obstacle for quantum computation in such systems. 
Further numerical studies show that this situation is not altered qualitatively
when the inversion symmetry is broken.

Moreover, we have introduced an appropriate measure of entanglement which 
takes explicitly into account the possibility of double occupancies.
This quantity allows to quantify the entanglement of the quantum state
during a gate operation and  
provides a necessary and sufficient condition for entangled states. 
Hence we expect this measure of 
entanglement to be useful in general in the study of quantum information
phenomena in systems such as  (real or artificial) diatomic molecules, or
other quantum-confined two-site structures.

%%%%%%%%%%%%%%%%%%%%%%%%%%%%%%%%%%%%%%%%%%%%%%%%%%%%%%%%%%%%%%%%%%%%%%

\acknowledgements{We thank Guido Burkard for useful discussions and
comments on this paper. J.~S. was supported by the Deutsche 
Forschungsgemeinschaft under Grant No. Schl 539/1-1 and acknowledges the
hospitality of the Institute for Theoretical Physics of Hannover University,
Germany, where this work was completed. D.~L. acknowledges partial
support from the Swiss National Science Foundation.
M acknowledges support from the National Science Foundation
under grant DMR-9714055.

%%%%%%%%%%%%%%%%%%%%%%%%%%%%%%%%%%%%%%%%%%%%%%%%%%%%%%%%%%%%%%%%%%%%%%

\appendix
\section{}
\label{appendix1}

Here we give the proof that $\eta(w)=0$ is indeed a suffcient condition
for $|w\rangle$ being a single Slater determinant state. 
The proof consists of two steps.

{\bf (i) Let w be purely real.}
Since $\eta(w)=0$ implies $\det w=0$ (cf. eq. (\ref{det})),
$w$ has at least one zero eigenvalue.
Because $w$ is real and antisymmetric its eigenvalues are purely imaginary
(if not zero) and occur in pairs of complex conjugates. Therefore, 
at least two of the four eigenvalues of $w$ are zero. 
It follows from standard arguments
(similar to those for real and symmetric matrices) that these two zero 
eigenvalues correspond to two real eigenvectors being orthogonal onto each
other. It follows that there is a real and orthogonal one-particle 
transformation $U$ so that, say, the first two rows and columns of the
resulting matrix $UwU^{T}$ are zero. Hence, the one-particle states with 
labels $a=1,2$ (in this new basis) are strictly empty, and the two electrons
occupy the remaining two states. Thus, $|w\rangle$ is clearly a single
Slater determinant.  

{\bf (ii) General case: w complex.}
By a one-particle transformation $\cal U$ with
\begin{equation} 
U={\rm diag}(e^{i\phi_{1}},e^{i\phi_{2}},e^{i\phi_{3}},e^{i\phi_{4}})
\end{equation}
one can adjust the phases in $w'=UwU^{T}$ in a manner that, say,
$w'_{12}$, $w'_{13}$, $w'_{14}$ are real. Denoting the real and imaginary
part of $w'$ by 
\begin{equation}
w'=u+iv
\end{equation}
it follows that $\det v=0$. Consider now the (unnormalized) states 
$|u\rangle$ and $|v\rangle$. If one of these states vanishes the assertion is 
already proved in (i), thus assume $|u\rangle\neq 0\neq|v\rangle$.
The condition $\eta(w)=0$ reads
\begin{equation}
\langle\tilde u|u\rangle-\langle\tilde v|v\rangle
+i\left(\langle\tilde u|v\rangle+\langle\tilde v|u\rangle\right)=0
\end{equation}
Since both terms in the imaginary part are equal by definition and
$\det v=0$ implies $\langle\tilde v|v\rangle=0$, it holds
\begin{equation}
\langle\tilde u|u\rangle=0\quad\Rightarrow\quad\det u=0
\end{equation}
and 
\begin{equation}
\langle\tilde u|v\rangle=\langle\tilde v|u\rangle=0\,.
\label{ortho}
\end{equation}
From (i) one concludes that both $|u\rangle$ and $|v\rangle$ are single
Slater determinants. Thus there are spinors $x^{1}$, $x^{2}$ and  
$y^{1}$, $y^{2}$ with
\begin{equation}
u_{ab}=\frac{1}{2}\left(x^{1}_{a}x^{2}_{b}-x^{1}_{b}x^{2}_{a}\right)\quad,\quad
v_{ab}=\frac{1}{2}\left(y^{1}_{a}y^{2}_{b}-y^{1}_{b}y^{2}_{a}\right)\,.
\end{equation}
Moreover, equation (\ref{ortho}) implies that
\begin{equation}
\varepsilon^{abcd}x^{1}_{a}x^{2}_{b}y^{1}_{c}y^{2}_{d}=0\,.
\end{equation}
Thus, the 4$\times$4-matrix having these four spinors as its rows or columns 
has a vanishing determinant. Therefore these spinors are linearly dependent.
Without loss of generality, consider the case
\begin{equation}
x^{1}=\alpha x^{2}+\beta y^{1}+\gamma y^{2}\,,
\end{equation}
where the complex coefficients $\beta$ and $\gamma$ are not both zero since
otherwise $u=0$. Let, again without loss of generality, $\beta$ be nonzero.
Then the spinors
\begin{equation}
z^{1}=\beta y^{1}+\gamma y^{2}\quad,\quad
z^{2}=x^{2}+\frac{i}{\beta}y^{2}
\end{equation}
solve the problem, i.e.
\begin{equation}
w'_{ab}=u_{ab}+iv_{ab}=
\frac{1}{2}\left(z^{1}_{a}z^{2}_{b}-z^{1}_{b}z^{2}_{a}\right)\,.
\end{equation}
$z^{1}$ and $z^{2}$ are both nonzero and not collinear to each other since
otherwise $w'=0$. Thus, up to an unimportant orthonormalisation, these two
spinors define one-particle states which allow to express $|w'\rangle$
(and consequently $|w\rangle$) as a single Slater determinant.

%%%%%%%%%%%%%%%%%%%%%%%%%%%%%%%%%%%%%%%%%%%%%%%%%%%%%%%%%%%%%%%%%%%%%%

\begin{figure}
\centerline{\includegraphics[width=8cm]{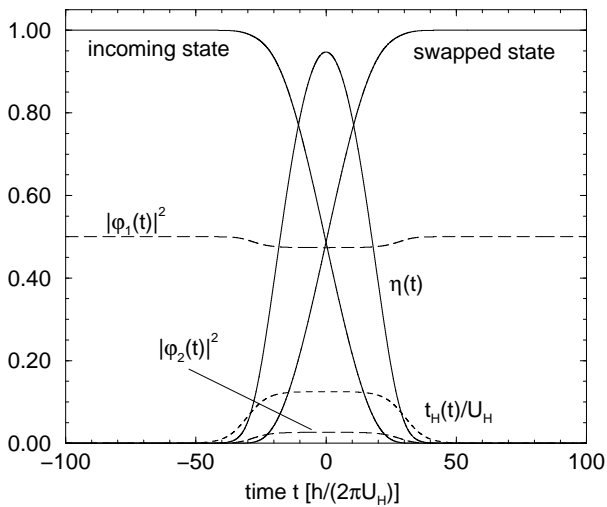}}
\caption{A swap process as a function of time. The 
tunneling amplitude $t_{H}(t)$ is plotted (in units of $U_{H}$) as a dotted 
line.
The square amplitude of the incoming state (\protect{\ref{instate}})
and the outgoing state (\protect{\ref{outstate}}) are
shown as thick lines. The square amplitudes of the singlets 
$|S_{1}\rangle$ and $|S_{2}\rangle$ are denoted by $|\varphi_{1}|^{2}$ and
$|\varphi_{2}|^{2}$, respectively, and plotted as long-dashed lines.
The measure of entanglement $\eta(t)$ is also shown.
\label{fig1}}
\end{figure}

\begin{figure}
\centerline{\includegraphics[width=8cm]{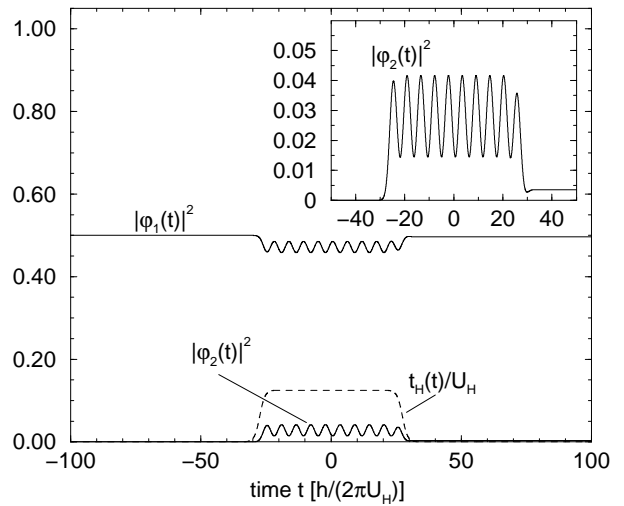}}
\caption{The square amplitudes of the singlet states $|S_{1}\rangle$ and 
$|S_{2}\rangle$ for the same situation as in figure \protect{\ref{fig1}}, but
with a four times smaller ramp time of only $\tau=\hbar/U_{H}$. 
The inset shows 
$|\varphi_{2}(t)|^{2}$ on a magnified scale. The dynamics of the system is
clearly in the non-adiabatic regime. 
\label{fig2}}
\end{figure}

\begin{figure}
\centerline{\includegraphics[width=8cm]{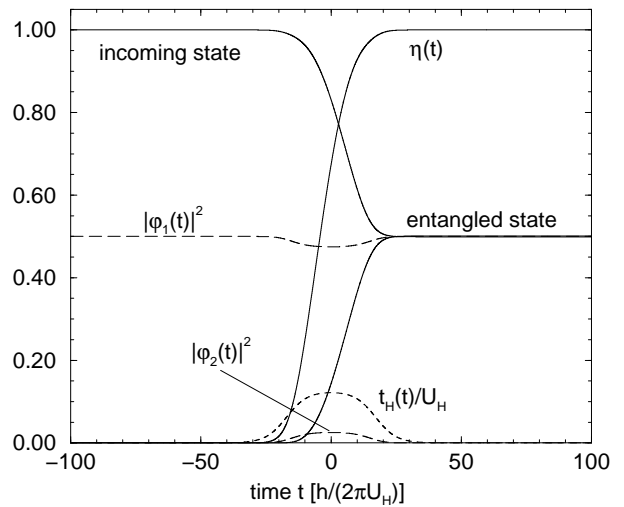}}
\caption{A square root of a swap, which is obtained from the situation of
figure \protect{\ref{fig1}} by halfing the pulse duration $T$. The probability
of double occupancies is again strongly suppressed after the tunneling pulse.
The resulting state is a fully entangled complex linear combination of 
$|S_{1}\rangle$ and $|T^{0}\rangle$, or, equivalently, of the 
incoming state (\protect{\ref{instate}}) and the outgoing state 
(\protect{\ref{outstate}}) of the full swap. The quantum mechanical weigths 
of the latter states are plotted as thick solid lines
\label{fig3}}
\end{figure}

\begin{figure}
\centerline{\includegraphics[width=8cm]{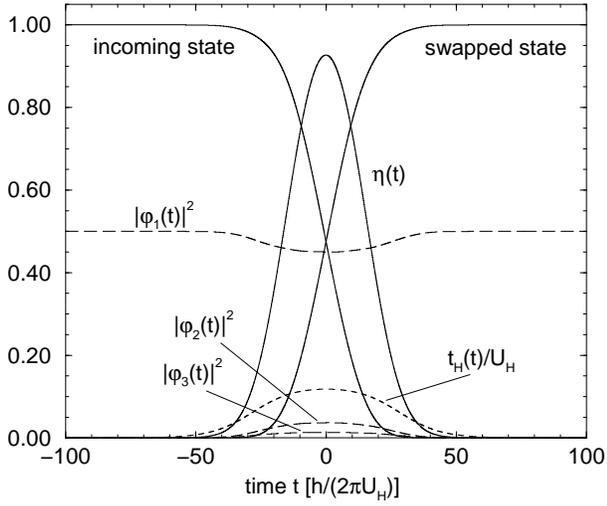}}
\caption{A swapping processes in the absence of inversion symmetry along
the axis connecting the dots. The square amplitudes of the singlet states
$|S_{i}\rangle$, $i\in\{1,2,3\}$, are denoted by $|\varphi_{i}|^{2}$.
The additional matrix elements entering the
Hamiltonian (\ref{3ham}) are $X=0.2U_{H}$ and $F=0.4U_{H}$.
The parameters of the tunneling pulse are
$\tau=8\hbar/U_{H}$ and $\Delta=U_{H}/8$ with a duration $T$ appropriate
for a single swapping. As a result, a clean swapping operation can be
performed also in the absence of inversion symmetry.
\label{fig4}}
\end{figure}

\end{document}